\documentclass[a4paper,showpacs,twocolumn,preprintnumbers,aps,amsmath,amssymb,nofootinbib,floatfix]{revtex4}
\usepackage{graphicx}
\usepackage{longtable}
\usepackage{float}
\usepackage{dcolumn}
\usepackage{bm}
\usepackage{appendix}
\usepackage{multirow}
\usepackage{color}

\newcommand{\Mpc}{\mathrm{Mpc}}

\begin{document}

\title{Bayesian analysis of inflationary features in Planck and SDSS data} 

\author{Micol Benetti\footnote{E-mail: micolbenetti@on.br}}

\author{Jailson S. Alcaniz\footnote{E-mail: alcaniz@on.br}}

\affiliation{Observat\'orio Nacional, 20921-400, 
Rio de Janeiro - RJ, Brazil}

\date{\today}

\begin{abstract}

We perform a Bayesian analysis to study possible features in the  primordial inflationary power spectrum of scalar perturbations. In particular, we analyse the possibility of detecting the imprint of these primordial features in the anisotropy temperature power spectrum of the Cosmic Microwave Background (CMB) and also in the matter power spectrum $P(k)$.  We use the most recent CMB data provided by the Planck Collaboration and $P(k)$ measurements from the eleventh data release of the Sloan Digital Sky Survey. We focus our analysis on a class of potentials whose features are localised at different intervals of angular scales, corresponding to multipoles in the ranges $10 < \ell < 60$ (Oscill-1) and  $150 < \ell < 300$ (Oscill-2). Our results show that one of the step-potentials (Oscill-1) provides a better fit to the CMB data than does the featureless $\Lambda$CDM scenario, with a {\textit{moderate}} Bayesian evidence in favor of the former. Adding the $P(k)$ data to the analysis weakens the evidence of the 
Oscill-1 potential relative to the standard model and strengthens the evidence  of this latter scenario with respect to the Oscill-2 model.

\end{abstract}

\pacs{98.80.-k, 98.80.Es, 98.80.Cq, 98.65.Dx}
\keywords{keywords}

\maketitle

\section{Introduction}

The inflationary paradigm offers an elegant theoretical framework in which the emergence of the primordial curvature perturbations can be understood. In the simplest inflationary scenarios a primordial scalar perturbation with nearly scale-invariant power spectrum is generated by a single minimally-coupled scalar field $\phi$ rolling down a smooth potential $V(\phi)$. This framework seems to agree with the most recent cosmic microwave background (CMB) data \cite{PLA-XX-2015, PLA-XIII-2015}, which show a preference for plateau-like\footnote{For plateau-like potentials  $V(\phi) \rightarrow 0$ as $\phi \rightarrow \infty$ (see, e.g., \cite{Starobinsky:1980te,Stewart:1994ts,Dvali:1998pa,Conlon:2005jm}).} over monomial potentials, as well as no compelling statistical evidence for a specific class of scenarios (we refer the reader to \cite{Ijjas:2013vea,Linde:2014nna,Guth:2013sya,Martin:2015dha,Brandenberger:2015kga} for different points of view of the current observational status of inflation).

Recently, several works have analysed inflationary models that account for localised features in the primordial power spectrum, showing in some cases a better fit to the data with respect to a smooth power-law spectrum \cite{Peiris:2003ff,Covi:2006ci,Hamann:2007pa,Mortonson:2009qv, Hazra:2010ve,Meerburg:2011gd, Benetti:2011rp,Benetti:2012wu, Benetti:2013cja, Hu:2014hra, Miranda:2013wxa}. 
Features in the primordial power spectrum can be generated following departures from the slow-roll approximation, which can happen in more general inflationary scenarios with a symmetry-breaking phase transition. Examples are the inflationary models with a step in the primordial potential, whose oscillation in the power spectrum of curvature perturbations is localised around the scale that crosses  the horizon at the time the phase transition occurs~\cite{Adams:2001vc,Hunt:2004vt}. 

In principle, this oscillation could be seen in several observables, such as  in the CMB maps and in the large-scale galaxy distribution, since the primordial scalar fluctuation is the seed for all the cosmic structures currently observed. It is therefore expected that data of the temperature anisotropy power spectrum as well as measurements of the matter power spectrum $P(k)$ from galaxy surveys may provide hints on the origin of these features. Originally, step-like inflationary potentials were postulated to study the ``glitch" appearing at the low-$\ell$ region of the CMB anisotropy  spectrum. The signatures of this class of models in the CMB temperature power spectrum and bispectrum~\cite{Benetti:2011rp, Benetti:2012wu, Benetti:2013cja, Meerburg:2011gd, Adshead:2011jq, Adshead:2011bw, Bartolo:2013exa, Cadavid:2015iya, Mooij:2015cxa, Fergusson:2014hya, Fergusson:2014tza, Miranda:2013wxa} and in the tensor spectrum \cite{Miranda:2014fwa, Miranda:2014wga} have been studied in detail using the current data and have shown a consistent improvement of  the $\Delta \chi^2$ values with 
respect to the featureless $\Lambda$CDM model.  

In this work we proceed a step further in this kind of analysis and employ a Bayesian statistical analysis to verify the predictions of a class of inflationary step-like potential models and discuss their observational viability in a wide range of scales, considering not only their observational consequences on the present CMB anisotropy temperature maps but also on the matter power spectrum $P(k)$. We 
work with two data sets: the second CMB data release of the Planck Collaboration~\cite{PLA-XX-2015, Aghanim:2015xee} and a combination of these CMB data with measurements of the matter power spectrum from  the Baryon Oscillation Spectroscopic Survey (BOSS) CMASS Data Release-$11$ sample of the Sloan Digital Sky Survey (DR11-SDSS) experiment (hereafter ``CMB+SDSS")~ \cite{Beutler:2013yhm}. We perform a Bayesian analysis using both the Metropolis-Hastings algorithm implemented in {\sc CosmoMC}~\cite{Lewis:2002ah} and the nested sampling algorithm of {\sc MultiNest}~\cite{Feroz:2008xx,Feroz:2007kg,Feroz:2013hea}. We find that at least one of the scenarios studied is able to provide a better fit to the CMB and CMB + $P(k)$ data than does the standard $\Lambda$CDM model.

This paper is organised as follows. Sec.~\ref{sec:model&method} reviews the class of inflationary models considered in this work. We also discuss observational data sets and priors used in the analysis as well as the Bayesian model selection method adopted. In Sec.~\ref{sec:results} we discuss the results and present a brief comparison with previous analysis. We end the paper by summarising the main results in Sec.~\ref{sec:conclusion}.

\section{MODEL AND METHOD}
\label{sec:model&method}

Relaxing the slow-roll condition leaves traces on the primordial power spectrum. For example, if the wavelengths cross the horizon during the fast-roll phase, one must expect deviations from the usual power-law power spectrum. The brief violation of the slow-roll condition can be shaped, in a single-field model, by adding a local feature, such as a step, 
to an otherwise flat potential. 
Following the formalism presented in Refs.~\cite{Adams:1996yd, Adams:1997de, Adams:2001vc}, we consider a model with a local feature added to a chaotic potential 
%
\begin{equation}
V(\phi) = \frac{1}{2}m^2\phi^2 \left[1+ c\tanh\left(\frac{\phi-b}{d}\right)\right] .
\label{eq:Vstep}
\end{equation}
The spectrum of primordial perturbations, resulting from the potential (\ref{eq:Vstep}), is found to be essentially a power-law with superimposed oscillations. These are centred on a value that depends on the parameter $b$, with amplitude set by $c$ and damping given by $d$. Asymptotically, the spectrum recovers the familiar $k^{n_s-1}$ form, typical of slow-roll inflationary models. 

In this work we adopt an analytical parametrisation for the scalar primordial spectrum resulting from Eq. (\ref{eq:Vstep}), as studied in \cite{Adshead:2011jq}:
\begin{equation} 
P_R(k) = \exp \left[ \ln P_0 (k) + \frac{A_f}{3} \frac {k\eta_f}{\sinh\ (\frac{k\eta_f}{x_d})} W'(k\eta_f) \right] \;, 
\label{eq:Adshead1} \\
\end{equation}
where
\begin{equation}
W'(x) = (-3+ \frac {9}{x^2})\cos 2x + (15-\frac{9}{x^2}) \frac{\sin 2x}{2x} \;,
\label{eq:Adshead2}
\end{equation}
 $P_0 (k) = A_s (\frac{k}{k_*})^ {n_s - 1}$ is the smooth spectrum with the standard power-law form, $A_f$ is the kinetic energy perturbation of the step, $\eta_f$ is the step crossing time in units of Mpc and $x_d$ is the dimensionless damping scale. The oscillating window function $W'(k\eta_f)$ is modulated by the decaying envelope ${k\eta_f}/{\sinh({k\eta_f}/{x_d})}$ which is set by the details of the step. It is worth mentioning that the approximate approach of  Eqs. (\ref{eq:Adshead1}) and  (\ref{eq:Adshead2}) allows to significantly save computing time with respect to the full numerical solution of the evolution equations for the potential of Eq. (\ref{eq:Vstep}), without a meaningful loss of accuracy \cite{Benetti:2013cja}.

We consider a ``vanilla" \newcommand{\LCDM}{\Lambda\mathrm{CDM}} model with the addition of features in the primordial spectrum, parametrised as in Eqs. (\ref{eq:Adshead1}) and  (\ref{eq:Adshead2}). In our analysis, we vary the usual cosmological parameters, namely, the physical baryon density, $\Omega_bh^2$, the physical cold dark matter density, $\Omega_ch^2$, the ratio between the sound horizon and the angular diameter distance at decoupling, $\theta$, the optical depth, $\tau$, the primordial scalar amplitude, $\mathcal A_s$, the primordial spectral index, $n_s$, and the additional $A_f$, $\eta_f$ and $x_d$ step parameters. We also vary the nuisance foreground parameters~\cite{Aghanim:2015xee} and consider purely adiabatic initial conditions. The sum of neutrino masses if fixed to $0.06$ eV, and we limit the analysis to scalar perturbations with $k_0=0.05$ $\rm{Mpc}^{-1}$. 

The posterior probabilities distributions of the parameters are generated using both the Metropolis-Hastings algorithm implemented in {\sc CosmoMC}~\cite{Lewis:2002ah} and the nested sampling algorithm of {\sc MultiNest}~\cite{Feroz:2008xx,Feroz:2007kg,Feroz:2013hea}. 
We use a modified version of the {\sc CAMB}~\cite{camb} code in order to compute the CMB anisotropies spectrum for different values of the parameters describing the step-like inflationary model, as in Eqs. (\ref{eq:Adshead1}) and  (\ref{eq:Adshead2}). 
The Gelman and Rubin criteria~\cite{Gelman:1992zz, An98stephenbrooks} is used to evaluate the convergence of the Monte Carlo Markov chain (MCMC) analysis, demanding that $R -1 \leq 0.02$.
In our Bayesian analysis we use the most accurate Importance Nested Sampling (INS)~\cite{Cameron:2013sm, Feroz:2013hea} instead of the vanilla Nested Sampling (NS), {requiring} INS Global Log-Evidence error $< 0.1$.

\begin{table}[tbp]
    \centering
    \begin{tabular}{|c|ll|}
        \hline
        & & \\
        Parameter   & Prior (Oscill-1)    &  Prior (Oscill-2)\\
        \hline  \hline
        & & \\
        $A_f$       & Uniform(0.5, 1)  &  Uniform(0, 0.2) \\
        & &\\
        $ \ln{\eta_f}$  & Uniform(7, 8)      & Uniform(6.5, 8)\\
        & & \\
        $\ln{x_d}$ & Uniform(0, 1)      & Uniform(3, 5) \\
        \hline  
    \end{tabular}
    \caption{\label{tab:priors} Priors on the features parameters of Eqs. (\ref{eq:Adshead1}) and  (\ref{eq:Adshead2}).}
\end{table}

\subsection{Priors}

As mentioned earlier, the class of models considered in this analysis is able to produce localised oscillations in the primordial power-law potential. The oscillation spot is set by the parameter $\eta_f$ in the Eqs. (\ref{eq:Adshead1}) and  (\ref{eq:Adshead2}) and depends on the scale where the wavelengths cross the horizon. 
We note that, for increasing values of the step-crossing time parameter, the oscillation is shifted to larger scales (lower multipoles), until it disappears completely from the TT spectrum for values of  $\ln(\eta/\rm{Mpc}) >10$.  At the same time, for values of $\ln(\eta/\rm{Mpc}) < 5$ the oscillation is shifted to the high-$\ell$ part of the spectrum, i.e., to scales without interest to the present study.

Previous works ~\cite{Planck:2013jfk,Benetti:2013cja} identified two different ranges of multipoles for these oscillations from  the data: the first one lies in the interval $10<\ell<60$ (hereafter Oscill-1) whereas the second one lies in the interval $150<\ell<300$ (hereafter Oscill-2). 
In particular, these oscillation ranges are distinguished by different priors on the $A_f$ parameter{{, i.e. $A_f > 0.5$ for Oscill-1 and $A_f < 0.5$ for the Oscill-2.}}
We run our preliminary test following the guidelines of the previous studies, i.e.,  assuming the same priors on the feature parameter $A_f$, and setting for the parameters $\ln (\eta_f/Mpc)$ and $\ln x_d$ the intervals [$5:10$] and [$-1:5$], respectively. Using the {\sc CosmoMC} we found the value where the step parameter posterior probability drops to zero, and we set the priors for our analysis  as shown in Table I.
{{In order to estimate the impact of the prior choice on the $A_f$ parameter, we also repeat the analysis using the relaxed priors $A_f~[0:1],~~\ln(\eta_f/Mpc)~[5.5:8.5],~~\ln x_d~[0:2]$.}}

It is important to highlight the central role played, in the {\sc CosmoMC} analysis, by the prior ranges choice, since the introduced $\eta_f$ and $xd$ step parameters show multimodal posterior probability distributions. The MCMC analysis, indeed, can fail to fully explore all peaks which contain significant probability, especially if the peaks are very narrow. The {\sc MultiNest} algorithm, on the contrary, do not shows the same analysis problem. For this reason we use the {\sc CosmoMC} code only for the preliminary parameters estimation and we present in this work only the {\sc MultiNest} analysis results.
%
%
\begin{table*}
\centering
\caption{$68\%$ confidence limits for the cosmological and step parameters. 
The first columns-block refer to a minimal $\Lambda$CDM model with a featureless spectrum, using binned high-$\ell$ TT data; 
the second columns-block show the constraint of the inflationary step-like model using Oscill-1 prior of Tab.(\ref{tab:priors}) and binned high-$\ell$ TT data; 
the last columns-block show the constraint of the inflationary step-like model using Oscill-2 prior of Tab.(\ref{tab:priors}) and unbinned high-$\ell$ TT data.
The $\Delta \chi^2_{best}$ and the $\ln \mathcal{B}_{MM'}$ of Oscill-2 analysis refers to the difference with respect to the $\Lambda$CDM analysis using the unbinned high-$\ell$ TT data. 
These results refer to the Bayesian analysis results of the {\sc MultiNest} code analysis.}
\label{tab:Tabel}
\begin{tabular}{|c|cc|cc|cc|}
\hline
\hline

\multicolumn{1}{|c|}{$ $}&
\multicolumn{2}{c|}{\textbf{$\Lambda$CDM model}}& 
\multicolumn{2}{c|}{\textbf{Oscillation-1}}&
\multicolumn{2}{c|}{\textbf{Oscillation-2}}
\\ 
Parameter	& TT+lowP & CMB+SDSS & TT+lowP & CMB+SDSS & TT+lowP & CMB+SDSS\\
\hline
$100\,\Omega_b h^2$ 	
& $2.225 \pm 0.02$ & $2.232 \pm 0.02$ 		
& $2.222 \pm 0.021$ & $2.229 \pm 0.02$		
& $2.225 \pm 0.021$ & $2.233 \pm 0.02$		
\\
$\Omega_{c} h^2$	
& $0.1194 \pm 0.0019$ & $0.1182 \pm 0.0017$	
& $0.1202 \pm 0.0019$ & $0.1188 \pm 0.0017$
& $0.1195 \pm 0.0019$ & $0.1183 \pm 0.0017$
\\
$100\, \theta$ 
& $1.04091 \pm 0.00043$ & $1.04104 \pm 0.00042$	
& $1.04108 \pm 0.00045 $ & $1.04097 \pm 0.00042$
& $1.04094 \pm 0.00043 $ & $1.04111 \pm 0.00040$
\\
$\tau$
& $0.083 \pm 0.008$& $0.084 \pm 0.008$	
& $0.083 \pm 0.008$ & $0.084 \pm 0.008$
& $0.083 \pm 0.008$ & $0.083 \pm 0.008$
\\
$n_s$ 
& $0.9664 \pm 0.0052$ & $0.9685 \pm 0.0050$	
& $0.9637 \pm 0.0052$ & $0.9660 \pm 0.0050$	 
& $0.9667 \pm 0.0053$ & $0.9685 \pm 0.0049$	
\\
$\ln 10^{10}A_s$  \footnotemark[1]
\footnotetext[1]{$k_0 = 0.05\,\Mpc^{-1}$.}
& $3.100 \pm 0.0162$ & $ 3.099 \pm 0.0166$
& $3.102 \pm 0.0164$ & $ 3.100 \pm 0.0166$
& $3.100 \pm 0.0163$ & $ 3.097 \pm 0.0170$
\\
$A_f$ 
& $-$ & $-$ 									
& $0.75 \pm 0.14$ & $0.75  \pm  0.13$			
& $0.037 \pm 0.027$ & $0.04  \pm  0.02$
\\
ln $\eta_f / Mpc$ 	 
& $-$ & $-$
& $7.18 \pm 0.10$ & $7.18  \pm 0.09$
& $7.316 \pm 0.39$ & $7.21  \pm 0.33$
\\
ln $x_d$ 		    
& $-$ & $-$
& $0.49 \pm 0.27$ & $0.49 \pm 0.27$
& $3.92 \pm 0.57$ & $4.0 \pm 0.5$
\\
$\sigma_8$   
& $0.835 \pm 0.009$  & $ 0.829 \pm 0.009$	
& $0.836 \pm 0.009$ & $ 0.831 \pm 0.009$
& $0.834 \pm 0.009$ & $ 0.828 \pm 0.009$
\\
$H_0 $  \footnotemark[2]
\footnotetext[2]{[km s$^{-1}$ Mpc$^{-1}$]}
& $67.48 \pm 0.83$ & $ 68.00 \pm 0.74$
& $67.10 \pm 0.85$ & $ 67.72 \pm 0.73$
& $67.42 \pm 0.85$ & $ 67.98 \pm 0.73$
\\
\hline
\hline
$\Delta \chi^2_{best}$         
& $-$ & $ - $	
& $5.8$ & $ 5$
& $1.7$ & $8.4$
\\
%
$\ln \mathcal{B}_{ij}$ \footnotemark[3]
\footnotetext[3]{The associated error is calculated with the simple error propagation formula, assuming that the two measurements are uncorrelated:  
$\sigma^2(\ln \mathcal{B}_{ij})= \sigma^2(\ln \mathcal{E}_{i})+\sigma^2(\ln \mathcal{E}_{j})$}
& $-$ & $-$
& $3.91 \pm 0.03$ & $1.82 \pm 0.03$
& $0.58 \pm 0.04$ & $-1.64 \pm 0.07$
\\
\hline
\end{tabular}
\end{table*} 
\subsection{Data sets}
We use the second release of Planck data \cite{PLA-XX-2015, Aghanim:2015xee} (hereafter TT+lowP), namely the high-$\ell$ Planck temperature data (in the range of $30< \ell <2508$) from the 100-,143-, and 217-GHz half-mission TT cross-spectra and the low-P data by the joint TT, EE, BB and TE likelihood (in the range of $2< \ell <29$). 
More precisely, the latter data come from the best-fit temperature map obtained by the commander component separation algorithm applied to Planck $30$-$857$ GHz data, jointly with the Wilkinson Microwave Anisotropy Probe (WMAP) $9$-year observations between  $23$ and $94$ GHz \cite{Bennett:2012zja} and the Haslam {\it{et al.}} $408$ MHz survey~\cite{Haslam:1982zz}; the E and B maps are obtained from the $70$ GHz maps using their $30$ and $353$ GHz maps as foreground templates.

We combine the CMB data with the matter power spectrum measurements from  the Baryon Oscillation Spectroscopic Survey (BOSS) CMASS Data Release-$11$ sample (covering the redshift range $z = 0.43 - 0.7$) of the Sloan Digital Sky Survey (DR11-SDSS) experiment (hereafter ``CMB+SDSS"). 
We use the data sets of \cite{Beutler:2013yhm} and publicly available in the SDSS Collaboration website (www.sdss3.org).

\begin{figure*}
	\centering
   \includegraphics[width=0.49\hsize]{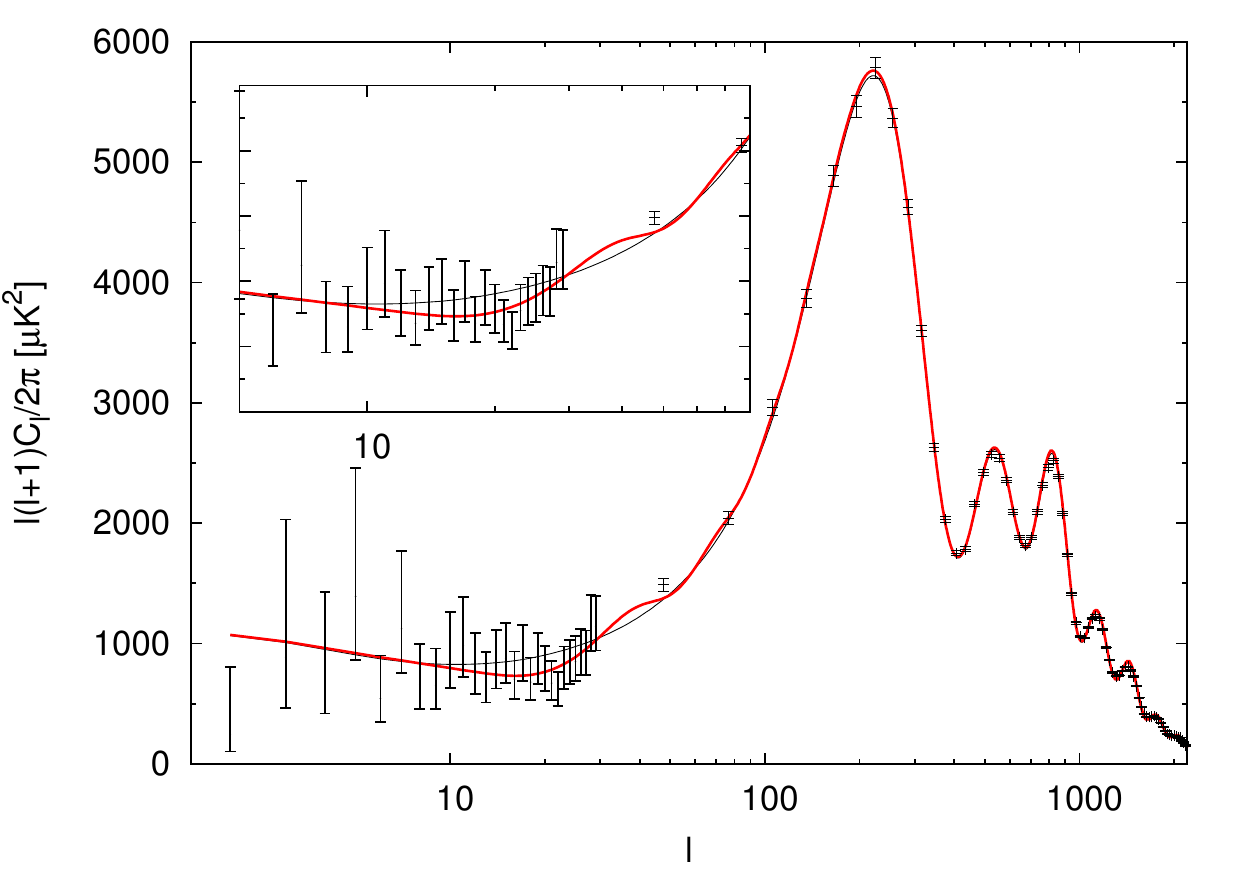}
      \includegraphics[width=0.49\hsize]{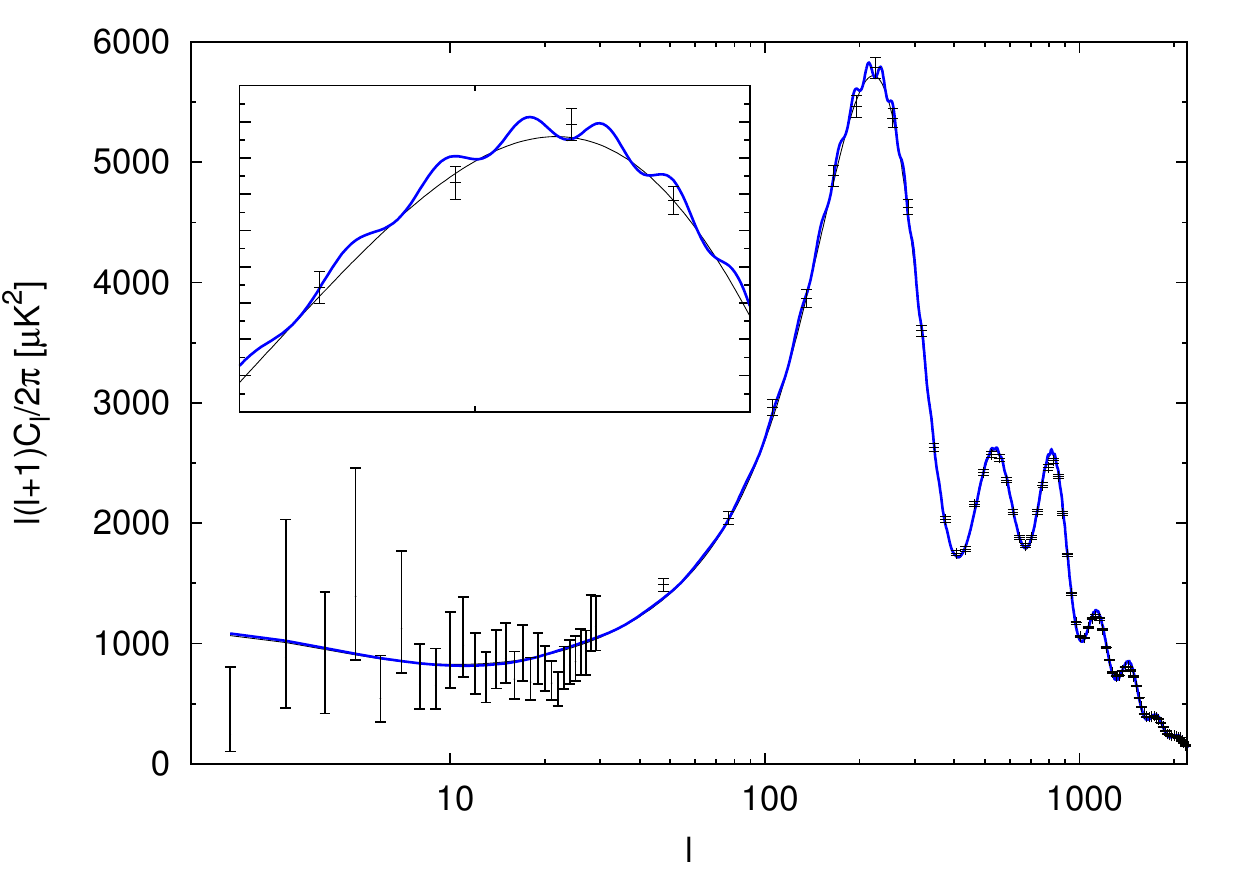}
       	\caption{Temperature power spectrum for the two inflationary step-like models best-fit values in comparison with the $\Lambda$CDM model best-fit (black line).
	{{Left}}: best-fit values for the Oscill-1 model using TT+lowP data. In the small box a zoom at $5<\ell<80$.
	{{Right}}: best-fit values for the Oscill-2 model using CMB+SDSS data. In the small box a zoom at $150<\ell<280$}
	\label{fig:TT}
\end{figure*}

\begin{figure}
	\centering
  \includegraphics[width=1\hsize]{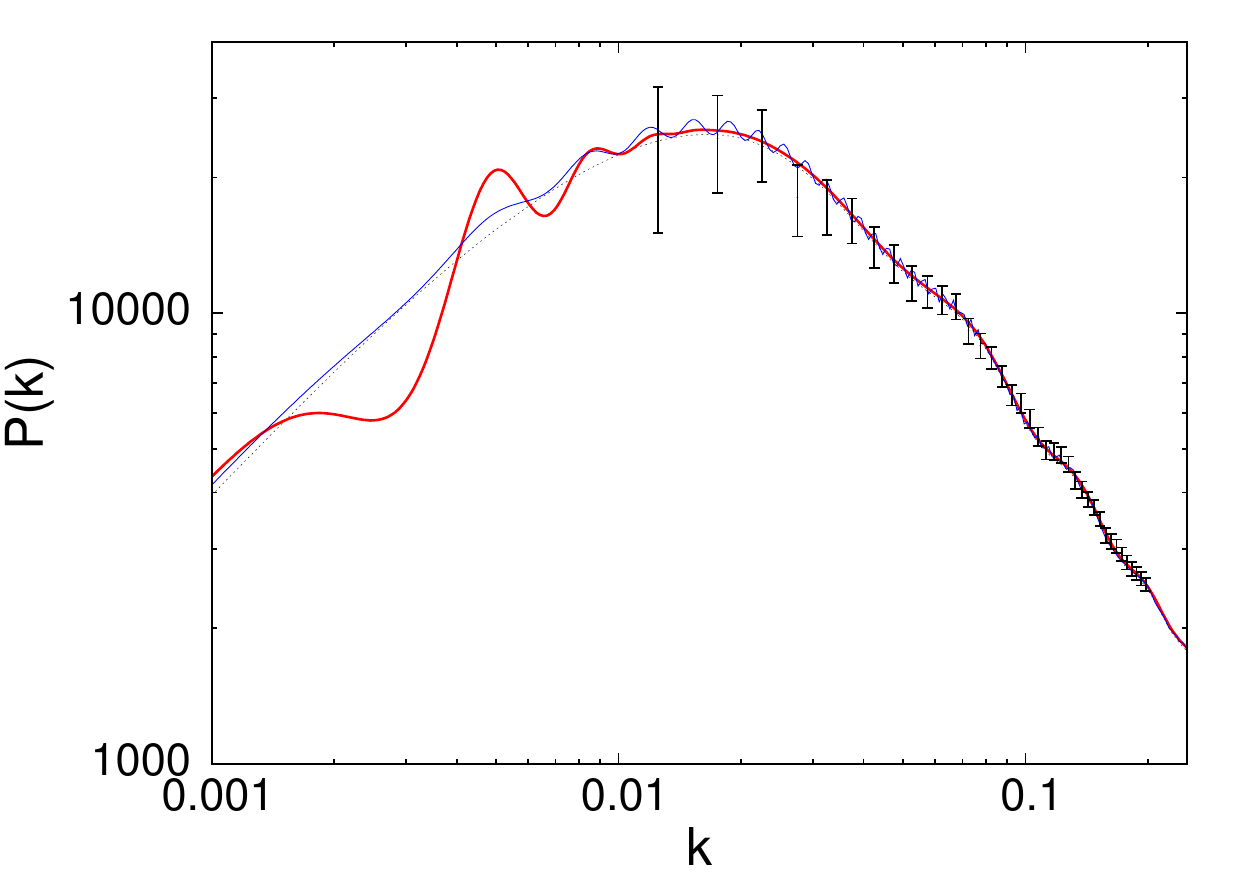}
       	\caption{Matter power spectrum for the best-fit $\Lambda$CDM model (dotted black line) and for the two step-like models: Oscill-1 best-fit values using TT+lowP data (red line), Oscill-2 best-fit values using CMB+SDSS data (blue line).}
	\label{fig:Pk}
\end{figure}

The best-fit CMB angular spectra and the matter power spectra for Oscill-1 and Oscill-2 models are shown in Fig.~\ref{fig:TT} and Fig. ~\ref{fig:Pk}. In order to obtain the results we use binned high-$\ell$ TT data for the analysis of the Oscill-1 model since this type of oscillation is placed in the low-$\ell$ region of the spectrum, which means that details on the high-$\ell$ part do not add any information. On the other hand, for the Oscill-2 model the oscillations are located at small scales and show higher frequence (see Fig.~\ref{fig:TT}, right panel). Therefore, for the analysis of this latter model we use the unbinned version of the high-$\ell$ TT data in order to maximise the sensitivity of sharp features that lie inside the single bin. 



  
%
%



\subsection{Statistical Model Selection}

In order to probe possible features in the primordial inflationary power spectrum, we perform a  Bayesian model comparison considering three models, namely, the standard $\Lambda$CDM scenario and the two models with localised oscillations in the power-law potential named earlier as Oscill-1 and Oscill-2. In this kind of  analysis the ``best'' model is the one that achieves the best compromise between quality of fit and 
predictivity. Indeed, while a model with more free parameters will always fit the data better (or at least as good as) a model with less parameters,  such added complexity ought to be avoided whenever a simpler model provides an adequate description of the observations. The Bayesian model comparison offers a formal way to evaluate whether the extra complexity of a model is required by the data, preferring the model that describes the data well over a large fraction of their prior volume.
%
In this section, we briefly review the model comparison and introduce our notation (we refer the reader to \cite{Liddle:2007fy,Trotta:2005ar,Parkinson:2006ku,thoven2016} for some recent applications of Bayesian model selection in cosmology).

Let us consider two competing models, $M_i$ and $M_j$, 
whose $n_j$ parameters (of model $M_j$) are common to $M_i$, which in turn has $n_i - n_j$ extra parameters. The posterior probability 
for the parameters vector $\theta$ (of length $n_i$) given the data $x$ under the model $M_i$
comes from Bayes' theorem:
\begin{equation}
p(\theta|x,M_i)=\frac{p(x|\theta,M_i)\pi(\theta|M_i)}{p(x|M_i)}\;,
\end{equation}
and similarly for $M_j$. The term $p(x|\theta,M_i)$ is the likelihood, while the $\pi(\theta|M_i)$ the prior probability distribution function.
The normalization constant in the denominator is called 
 \textit{evidence} $\mathcal{E}$, and is the marginal likelihood for the model $M_i$:
\begin{equation}
p(x|M_i) \equiv \mathcal{E}_{M_i} =\int d\theta p(x|\theta, M_i) \pi(\theta|M_i)\;.
\end{equation}
%
The posterior probability of the $M_i$ model given the data is written as
\begin{equation}
p(M_i|x) \propto \mathcal{E}_{M_i} \pi(M_i)\;.
\end{equation}
%
Assuming no {\it{a priori}} preference about any model ($\pi(M_j) = \pi (M_i)$), the ratio of the posterior probabilities of the two models (the so-called \textit{Bayes Factor}) is given by
\begin{equation}
{\mathcal{B}_{ij}}= \frac{\mathcal{E}_{M_{i}}}{\mathcal{E}_{M_{j}}}\;.
\end{equation}
The more complex model $M_i$ will (if $M_j$ is nested) inevitably lead to a higher 
(or at least equal)
likelihood,
but the evidence will favor the simplest model if the fit is nearly as good, through
the smaller prior volume.
We assume uniform (and hence separable) priors in each parameter, such that 
$\pi(\theta|M_i) = (\Delta \theta_1 ~. . .~\Delta\theta_{n_i})^{-1}$ and
\begin{equation}
\mathcal{B}_{ij}= \frac{ \int d\theta p(x|\theta,M_i)}{ \int d\theta' p(x|\theta',M_j)}
\frac{(\Delta \theta_1 ~. . .~\Delta\theta_{n_{i}})}{(\Delta \theta'_1 ~. . .~\Delta\theta'_{n_{j}})}
\end{equation}
%
%
\begin{figure*}[!]
	\centering
	\includegraphics[width=5.5in, height=4.95in]{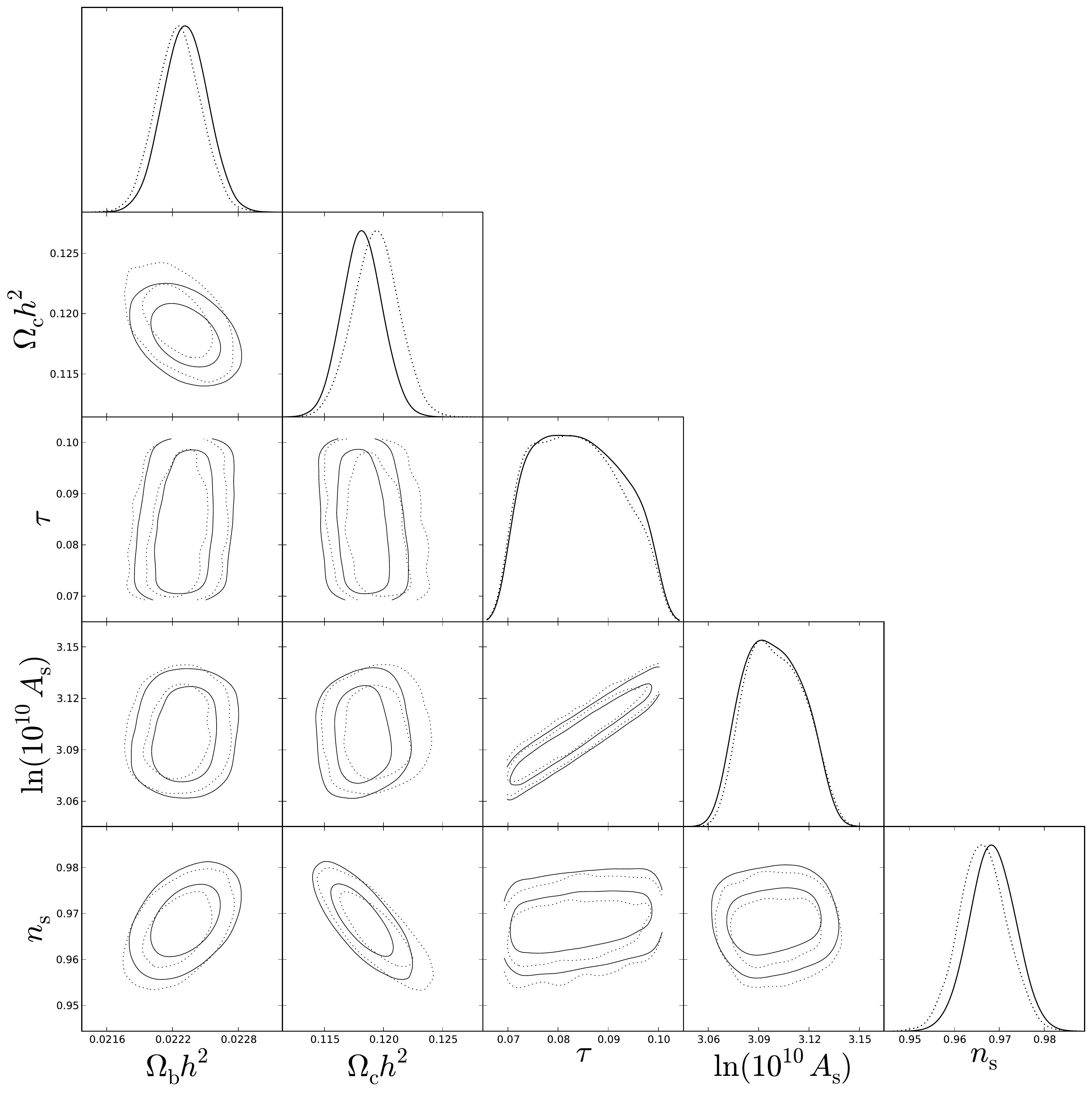}
	\caption{ { $68\%$ and $95\%$ confidence regions on $\Lambda$CDM model for Planck TT+lowP data (black dotted line) and CMB+SDSS data (black solid line). The numerical results of these analyses are reported in the first columns-block of Tab. \ref{tab:Tabel}.} }
	\label{fig:LCDM_triangle}
\end{figure*}
%
%
%
In order to rank the models of interest, we adopted the following scale to interpret the values of $\ln{\mathcal{B}_{ij}}$ in terms of the strength of the evidence of a chosen reference model ($M_j$): $ \ln{\mathcal{B}_{ij}} = 0 - 1 $ , $\ln{\mathcal{B}_{ij}}  =1 - 2.5$ , $ \ln{\mathcal{B}_{ij}}  =2.5 - 5$, and $ \ln{\mathcal{B}_{ij}}  >5 $ indicate, respectively, an {\textit{inconclusive}}, {\textit{weak}}, {\textit{moderate}} and {\textit{strong}} preference of the model $M_i$ with respect to the model $M_j$. 
Note that  
negative values of $\ln{\mathcal{B}_{ij}}$ means support in favour of the model $M_j$.
We refer to~\cite{Trotta:2005ar} for a more complete discussion about this scale, that
is a revised and more conservative version of the so-called \emph{Jeffreys' scale}~\citep{jeffreys}. 

\begin{figure}
\centering 
\includegraphics[width=1.05\hsize]{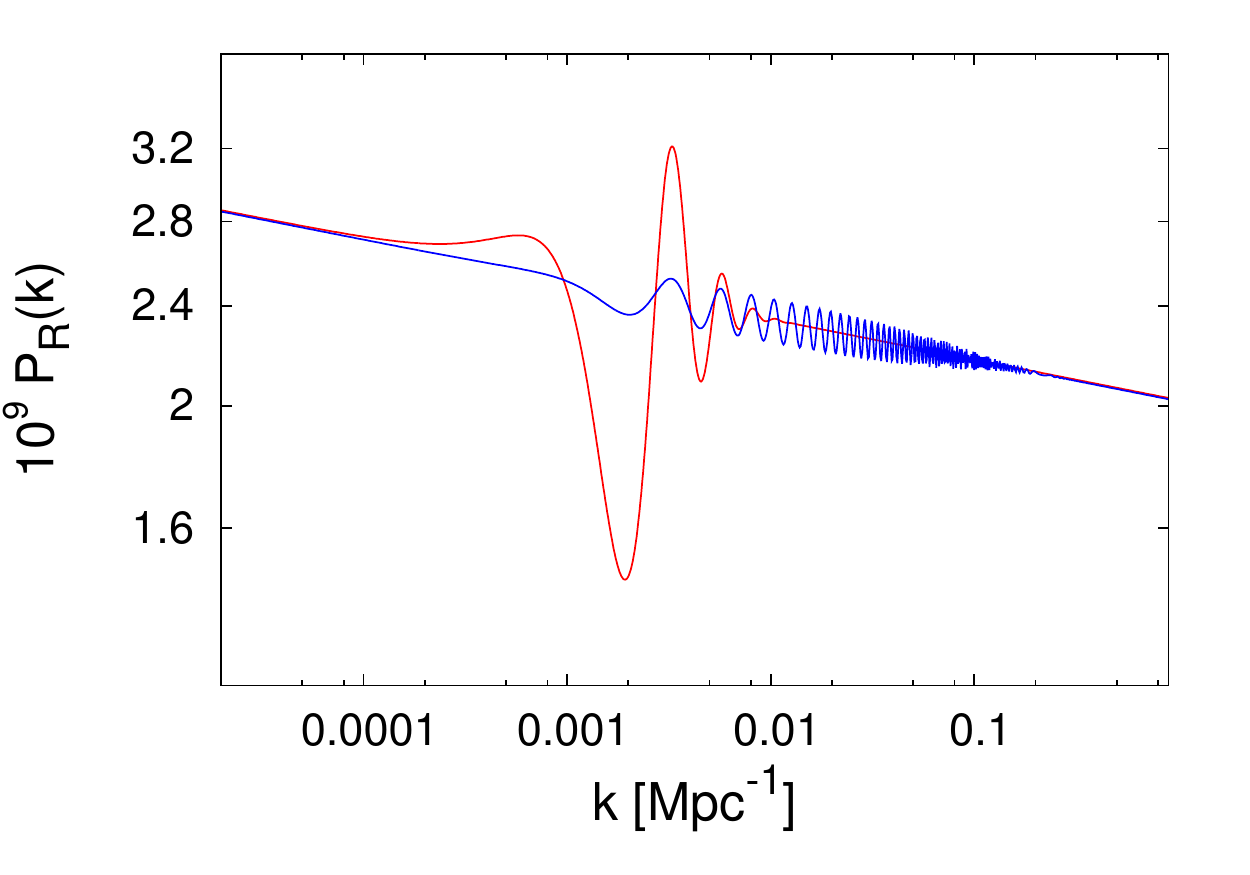}
\caption{Primordial power spectrum for the two step models: Oscill-1 (red line) with $A_f=0.75$, $\ln(\eta/Mpc)=7.18$ and $\ln x_d=0.49$ and Oscill-2 (blue line) with $A_f=0.04$, $\ln(\eta/Mpc)=7.21$, and $\ln x_d=4$ }
\label{fig:PR}
\end{figure}

\begin{figure*}
	\centering
	\includegraphics[width=0.77\hsize]{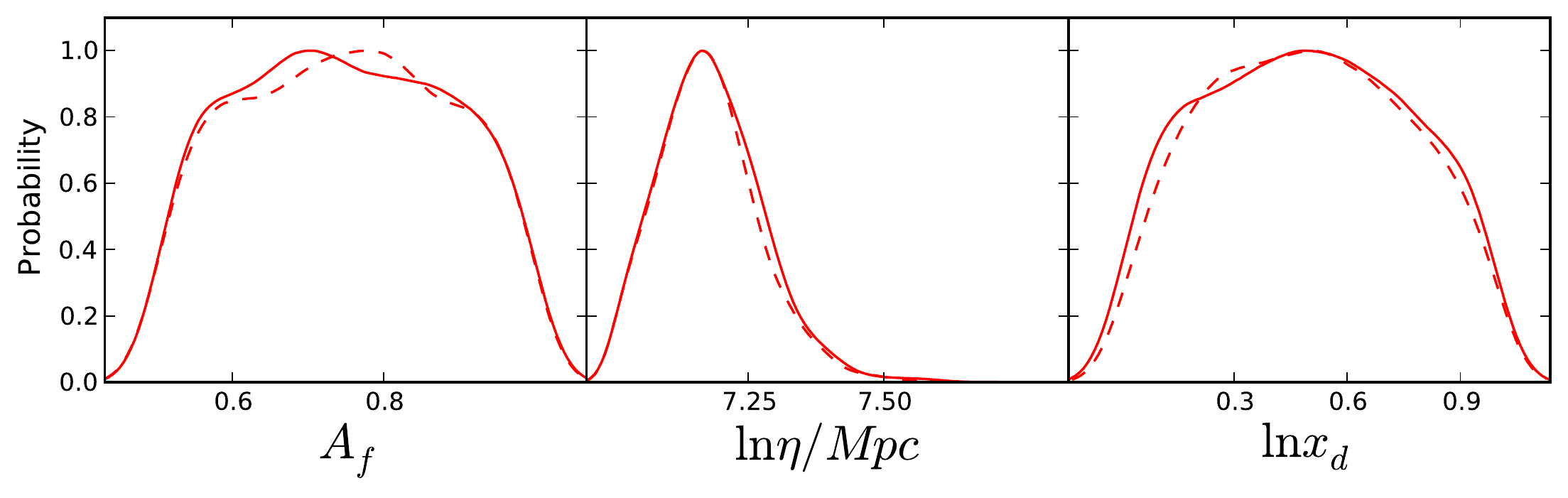}
	\includegraphics[width=0.77\hsize]{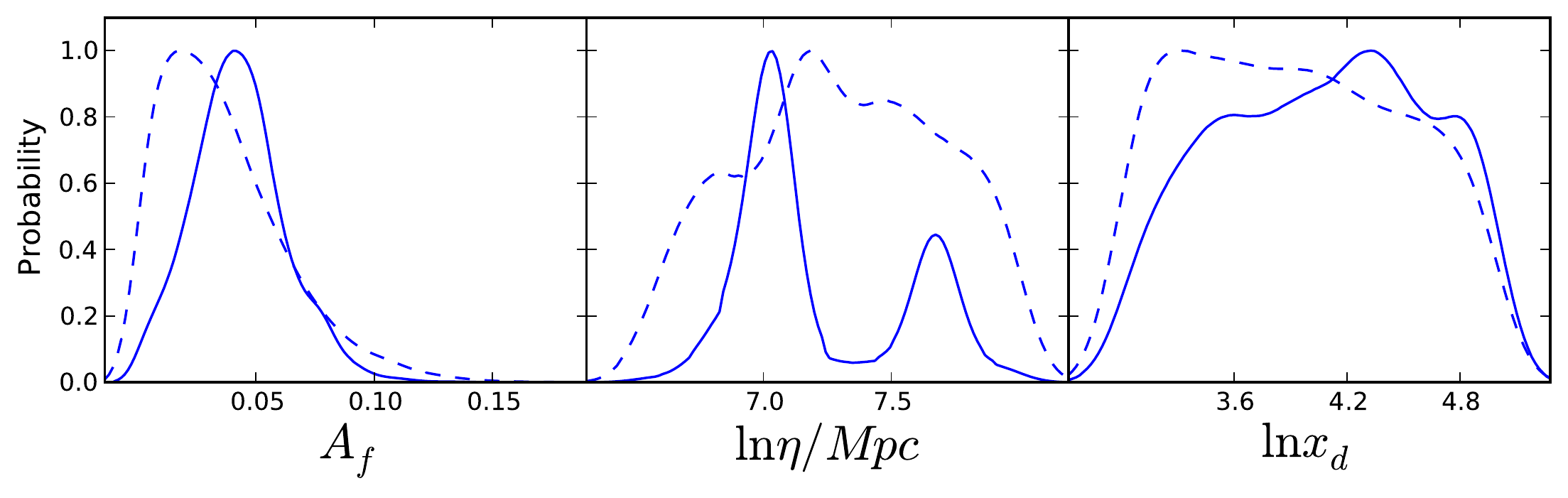}
	\caption{One-dimensional posterior probability densities. Top: Oscill-1 step parameters analysis reported in Tab.\ref{tab:Tabel} central columns-block, using TT+lowP data (red dashed line) and CMB+SDSS data (red solid line). Bottom: Oscill-2 step parameters analysis reported in the last columns-block of Tab.\ref{tab:Tabel} using TT+lowP data (blue dashed line) and CMB+SDSS data (blue line).}
	\label{fig:Oscill_posterior}
\end{figure*}
\begin{figure}
	\centering
	\includegraphics[width=1.05\hsize]{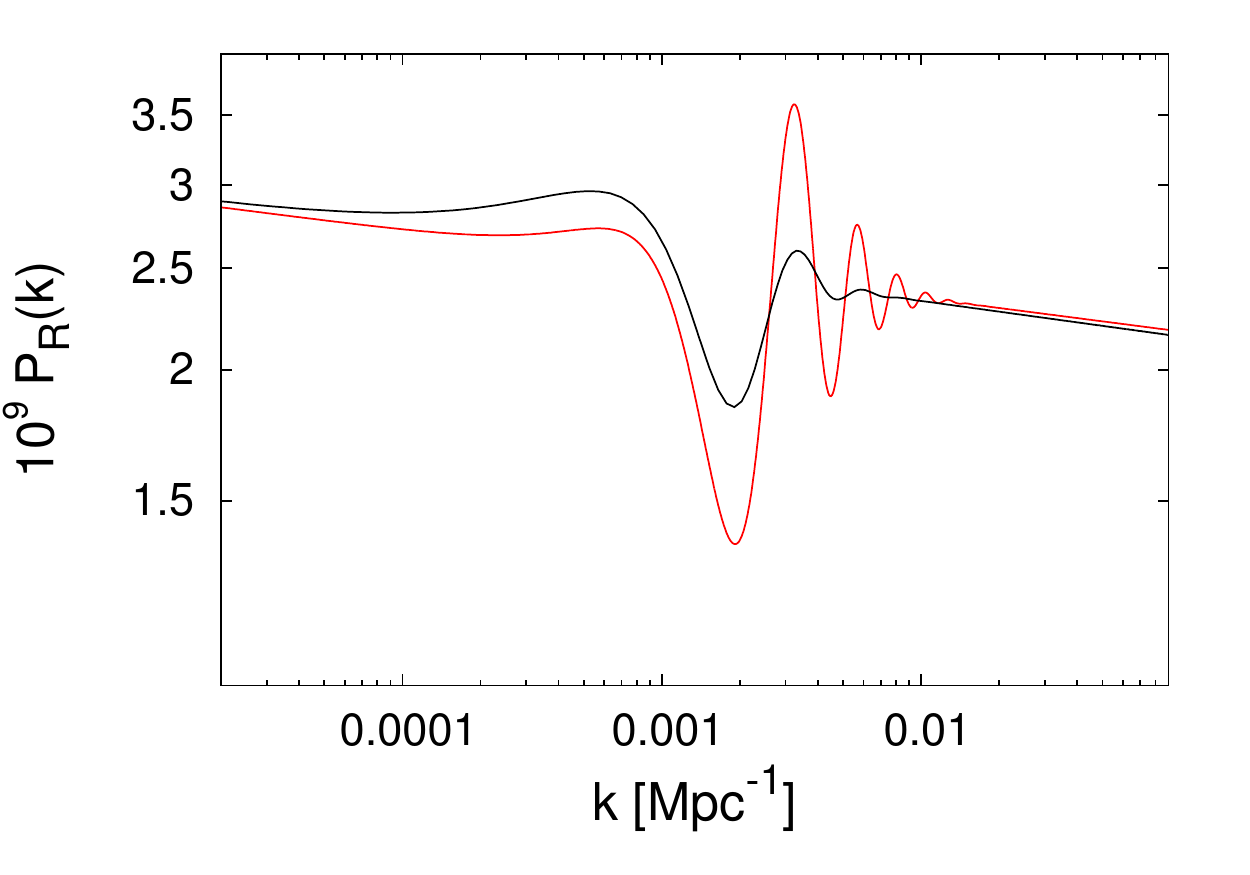}
	\caption{Primordial power spectrum for the Oscill-1 best-fit model (red line) with $A_f=0.728$, $\ln(\eta/Mpc)=7.20$ and $\ln x_d=0.75$, and for the parameterisation discussed in the Planck-2015 analysis~\cite{PLA-XX-2015} (black line) with the best-fit values $A_f=0.347$, $\ln(\eta/Mpc)=-3.10$, and $\ln x_d=0.342$.}
	\label{fig:PR_bestfit_confronto}
\end{figure}
\section{RESULTS}
\label{sec:results}

The main quantitative results of our analysis are shown in Tab.~\ref{tab:Tabel}.  
Firstly, we assume the minimal $\Lambda$CDM model and use the CMB and CMB+SDSS data sets discussed earlier. 
{From Fig.~\ref{fig:LCDM_triangle} one can see that the addition of the galaxy data (solid line) shows a preference for lower values of $\Omega_ch^2$ and for higher values of $n_s$, with respect to the analysis using only CMB (TT+lowP) data.
From Tab.~\ref{tab:Tabel} we also note a slightly improvement of the constraints on the Hubble parameter $H_0$ and a preference for higher value of $\sigma_8$.}
A good concordance between the results using binned and unbinned high-$\ell$ TT data is verified. For brevity, however, we report in Tab. \ref{tab:Tabel} only the $\Lambda$CDM analysis using the binned high-$\ell$ TT data. It is also worth mentioning that the values of $\Delta \chi^2_{best}$ and $\ln \mathcal{B}_{ij}$ for the Oscill-2 model are obtained with respect to the $\Lambda$CDM analysis using the unbinned high-$\ell$ TT data. 

For the central mean values given in Tab. \ref{tab:Tabel}, we show in Fig.~\ref{fig:PR} the primordial scalar power spectrum for the two step models.  We note that the two primordial oscillations start around the same scale (approximately the same value of $\eta_f$) but show very different amplitude and damping properties, which in turn produce very different features. The primordial power spectrum of the Oscill-1 model (red line) has a high amplitude and damping parameters values, and produces features at the scale interval $10^{-3} \lesssim k \lesssim 10^{-2}$. In principle, this makes possible to detect them using only CMB data, since the current LSS data cover scales of $k \gtrsim 10^{-2}$ \cite{Beutler:2013yhm}.

In Fig.~\ref{fig:Oscill_posterior} we show the posterior probability distribution for the step parameters values. As expected, the addition of the large-scale structure data improves the constraints on these parameters, mainly those related to the Oscill-2 model whose features extends to lower scales (see Fig.~{\ref{fig:PR}}). In comparison with the TT+lowP constraints (blue dashed line), we note the tigher constraints on the frequency parameter $\ln{\eta_f}/Mpc$, which for the CMB+SDSS data shows a bimodal distribution. 

Table \ref{tab:Tabel} also shows that the constraints on the usual cosmological model parameters are not significantly affected by the presence of primordial features. Moreover, in the case of the Oscill-1 model, our bounds on the step parameters are consistent with previous analysis~\cite{Benetti:2013cja} using the first Planck release (2013) 
with the only exception of the amplitude parameter $A_f$, which now prefers lower values.  On the other hand, our results for the TT+lowP data also shows a {\it{moderate}} evidence ($\ln{\mathcal{B}_{ij}}  = 3.91 \pm 0.03$) in favor of the Oscill-1 model with respect to the $\Lambda$CDM scenario.
If we relax the step-parameter priors to 
\begin{equation}
A_f~[0:1],~~\ln(\eta_f/Mpc)~[5.5:8.5],~~\ln x_d~[0:2]\;,
\end{equation}
the {\it{moderate}} evidence of Oscill-1 becomes {\it{weak}}, and the distribution of probability shows a weak secondary peak, which extends the explored parameter space. 
Our results seem to be in disagreement with those of the Planck Collaboration~\cite{PLA-XX-2015}, where a featureless primordial potential is preferable over the step-like models. 
Actually, the constrained oscillation in Ref.~\cite{PLA-XX-2015} 
refers to a different parametrization model \cite{Miranda:2013wxa} and shows different step parameters values, whose overall effect is to produce smaller and deeper oscillations with respect to the 
results of this work,  as shown in Fig.~\ref{fig:PR_bestfit_confronto}. 
At the same time, the parametrization used in this work produce a different behaviour in the scales immediately before where the oscillation occurs. 

For the Oscill-2 model, our analysis shows an a {\it{inconclusive}} evidence ($\ln{\mathcal{B}_{ij}}  = 0.58 \pm 0.04$) relative to the $\Lambda$CDM scenario using the TT+lowP dataset. 
We also observe that the addition of $P(k)$ data weakens the evidence of Oscill-1 model relative to the standard cosmology and strengthens the evidence  of this latter with respect to the Oscill-2 model.
%

\section{CONCLUSION}
\label{sec:conclusion}
We have performed a Bayesian model selection statistics to compare the observational viability of a class of inflationary models with step-like features in the inflaton potential and the $\Lambda$CDM cosmology using the most up-to-date CMB and LSS datasets. The step-like inflationary potentials studied are able to produce features in the primordial scalar power spectra, inducing an oscillation in the anisotropy power spectrum with magnitude, extent and position which depend on three step parameters. We have considered two types of models beyond the minimal $\Lambda$CDM model: Oscill-1 model, whose features lie in the multipole range $10 < \ell < 60$, and the Oscill-2 model, which produces features at multiples $150 < \ell < 300$ (see Fig.~\ref{fig:TT}).

In order to perform our analysis, we have used an approximate form of the power spectrum, as given in Eqs. (\ref{eq:Adshead1}) and  (\ref{eq:Adshead2}), and two data sets: the most recent data release of the Planck Collaboration (TT+ lowP) and this CMB data set added to the $P(k)$ measurements from the DR11 of the SDSS Collaboration. For the $\Lambda$CDM model, our analysis shows a good concordance between the results using the CMB data only and the extended dataset (CMB+SDSS). As the main result of this analysis, we have shown that the Oscill-1 model provides a better fit to current CMB data than does the standard $\Lambda$CDM model, with a {\it{moderate}} Bayesian evidence in favor of the step-like potential (Oscill-1). {Such result, however, seems to be in disagreement with the one reported by the Planck Collaboration, in which a featureless primordial potential is preferable over the step-like models.} When the extended data set (CMB+SDSS) is considered, the evidence of the Oscill-1 model relative to the standard cosmology becomes {\it{weak}}, just as the Bayesian evidence of the $\Lambda$CDM cosmology with respect to the Oscill-2 model.  

Finally, as shown in Fig.~\ref{fig:TT}, our best-fit scenario (Oscill-1 model) deviates significantly from the standard model only at very low values of $k$. We, therefore, expect to verify the reality of these features in the CMB and matter power spectra with data from the future release of the Planck collaboration, as well as from the next generation of very deep galaxy surveys like, for example, the Dark Energy Spectroscopic Instrument (DESI)~\cite{desi} and the Javalambre Physics of the Accelerating Universe Astrophysical Survey (J-PAS)~\cite{jpas}.

\section{Acknowledgments}
MB acknowledges financial support from the Rio de Janeiro Research Foundation (Post-doc {\textit{Nota 10}}). 
JSA is supported by CNPq, INEspa\c{c}o and FAPERJ. 
The authors thanks S\'ergio Fontes, Antonio Fran{\c{c}}a and Eduardo Matera for their assistance and 
acknowledge the use of CosmoMC~\cite{camb}
and the Multinest codes~\cite{Feroz:2008xx,Feroz:2007kg,Feroz:2013hea}. 
We are also grateful to Florian Beutler for helpful discussions and to the anonymous referee for the good comments and recommendations.

\end{document}